\documentclass[useAMS,usenatbib]{mn2e}
	\usepackage{graphicx}
    \usepackage{rotating}
    \usepackage{booktabs}
    \usepackage{lscape}
    \usepackage{amsmath}
    \usepackage[hang,small,labelfont=,textfont=]{caption}
\title[Very low-mass black hole - host galaxy relation]
  {The black hole - host galaxy relation for very low-mass quasars.}
\author[J.Sanghvi et al.]
  {J.~Sanghvi,$^1$
  J.K.~Kotilainen,$^2$ R.~Falomo,$^3$ R.~Decarli,$^4$
  K.~Karhunen,$^1$ M.~Uslenghi,$^5$\\
  $^1$Tuorla Observatory, University of Turku, V{\"a}is{\"a}l{\"a}ntie 20, 21500 Piikki{\"o}, Finland \\
  $^2$Finnish Centre for Astronomy with ESO (FINCA), University of Turku, V{\"a}is{\"a}l{\"a}ntie 20, 21500 Piikki{\"o}, Finland \\
  $^3$Osservatorio Astronomico di Padova, INAF, Vicolo dell' Osservatorio 5, 35122 Padova, Italy \\
  $^4$Max-Planck-Institut f{\"u}r Astronomie, K{\"o}nigstuhl 17, 69117 Heidelberg, Germany \\
  $^5$INAF-IASF - via E. Bassini 15, I-20133 Milano, Italy \\}
\date{Accepted for publication on 03 September 2014}

\pagerange{\pageref{firstpage}--\pageref{lastpage}} \pubyear{2014}

\def\LaTeX{L\kern-.36em\raise.3ex\hbox{a}\kern-.15em
    T\kern-.1667em\lower.7ex\hbox{E}\kern-.125emX}

\begin{document}

\label{firstpage}

\maketitle

\begin{abstract}
Recently, the relation between the masses of the black hole ($M_{BH}$) and the host galaxy ($M_{host}$) in quasars has been probed down to the parameter space of $M_{BH}\sim10^8 M_\odot$ and $M_{host}\sim10^{11} M_\odot$ at \textit{z} $<$ 0.5. In this study, we have investigated the $M_{BH}$ - $M_{host}$ log-linear relation for a sample of 37 quasars with low black hole masses ($10^7 M_\odot < M_{BH} < 10^{8.3} M_\odot$) at 0.5 $<$ \textit{z} $<$ 1.0. The black hole masses were derived using virial mass estimates from SDSS optical spectra. For 25 quasars, we detected the presence of the host galaxy from deep near-infrared \textit{H}-band imaging, whereas upper limits for the host galaxy luminosity (mass) were estimated for the 12 unresolved quasars. We combined our previous studies with the results from this work to create a sample of 89 quasars at \textit{z} $<$ 1.0 having a large range of black hole masses ($10^7 M_\odot < M_{BH} < 10^{10} M_\odot$) and host galaxy masses ($10^{10} M_\odot < M_{host} < 10^{13} M_\odot$). Most of the quasars at the low mass end lie below the extrapolation of the local relation. This apparent break in the linearity of the entire sample is due to increasing fraction of disc-dominated host galaxies in the low-mass quasars. After correcting for the disc component, and considering only the bulge component, the bilinear regression for the entire quasar sample holds over 3.5 dex in both the black hole mass and the bulge mass, and is in very good agreement with the local relation. We advocate secular evolution of discs of galaxies being responsible for the relatively strong disc domination.
\end{abstract}

\begin{keywords}
galaxies: active -- galaxies: bulges -- galaxies: nuclei -- quasars: general -- quasars: supermassive black holes
\end{keywords}

\section{Introduction}
Supermassive black holes (SMBHs) ubiquitously reside in the centres of massive galaxies \citep[e.g.][]{kr95,r98}. However, it is still debated as to what is the mechanism that drives the formation of the black hole and how the black hole is responsible for shaping the evolution of its host galaxy. A wealth of observations have shown that there is a very tight correlation between the mass of the SMBH, $M_{BH}$ and the large scale host galaxy properties, such as the stellar velocity dispersion $\sigma_*$, the luminosity of the host galaxy $L_{host}$ and the mass of the host galaxy $M_{host}$ \citep[e.g.][]{kr95,fm00,geb00}. This indicates a strong link between the evolution of the host galaxy and its central black hole (e.g. \citet{h07,m07,shankar09} but see for e.g. \citet{peng07,janke11} for contradicting views). In this context, accretion plays an important role by allowing the growth of BHs and thus causing the gas lying in the outskirts of the host galaxies to cool through feedback processes. This process quenches star formation. Consequently, galaxy mergers may also cause gravitationally induced dynamical instabilities, triggering bursts of star formation and gas inflows, thus fueling the BH activity \citep{kauff00,dimatteo05,can07,ben08,cis11}. 

Estimating the $M_{BH}$ of galaxies beyond the local Universe is challenging, because the radius of influence of the BHs can be resolved only for the most nearby galaxies, whereas more distant sources require indirect tracers of $M_{BH}$. Typically, the only available indirect tracer of $M_{BH}$ is only applicable for Type-1 AGN, based on a virial estimate of the velocity and size of the broad line region (BLR) from the broad emission-line widths and continuum luminosity \citep{peter00,vester02,vester06,dec10a}. This method has allowed to estimate the $M_{BH}$ from single-epoch spectra in $\sim$10$^5$ quasars up to \textit{z} $\sim$ 5 from SDSS spectra \citep{shen11} and in additional quasars up to \textit{z} $\sim$ 6 \citep{willott03,kurk07,willott10,derosa11}. On the other hand, the host galaxy properties can be more easily derived from analysis of deep imaging observations. 

In \citet{dec10a} and \citet{dec10b} we studied an extensive sample of quasars at 0 $<$ z $<$ 3. Of these, 64 quasars are at \textit{z} $<$ 1, and most of these quasars have $M_{BH}\geq10^9$M$_\odot$. More recently, in \citet{decarli2012low} we studied 25 quasars, most of them lying within $10^8<M_{BH}(M_\odot)<10^9$ at \textit{z}$<$0.5, and found a large fraction of these quasars to have disc-dominated host galaxies. In \citet{decarli2012low} we showed that the linear $M_{BH}$ - $M_{host}$ relation holds in quasars over a range of 2 dex in $M_{BH}$, from $10^8M_\odot$ to $10^{10}M_\odot$. For quiescent galaxies and low-luminosity AGN at low-\textit{z}, $M_{BH}$ - $M_{host}$ relation has been explored down to $M_{BH}\sim10^5 M_\odot$  \citep{gh07,t08,j11}. However, little is known about the black hole mass - bulge mass relation for very low-mass quasars. This study extends the study of \citet{decarli2012low} by investigating the $M_{BH}$ - $M_{host}$ mass relation down to $M_{BH}\sim10^7M_\odot$ and up to \textit{z} $<$ 1.0. 

The outline of the paper is as follows. In Section 2 we discuss how the black hole masses were estimated using the virial theorem to select the samples. In section 3 we discuss the observations, data reduction and analysis. Section 4 presents the results and discussion. In section 5, we present the conclusions of this work.

All the work in this paper has been performed assuming the following cosmological parameters: $H_0$=70 km s$^{-1}$ Mpc$^{-1}$, $\Omega_{\Lambda}$=0.7 and $\Omega_m$=0.3 .

\section{Sample selection and Virial Black hole mass estimation}

\begin{figure}
\includegraphics[height=8.5cm,angle=0]{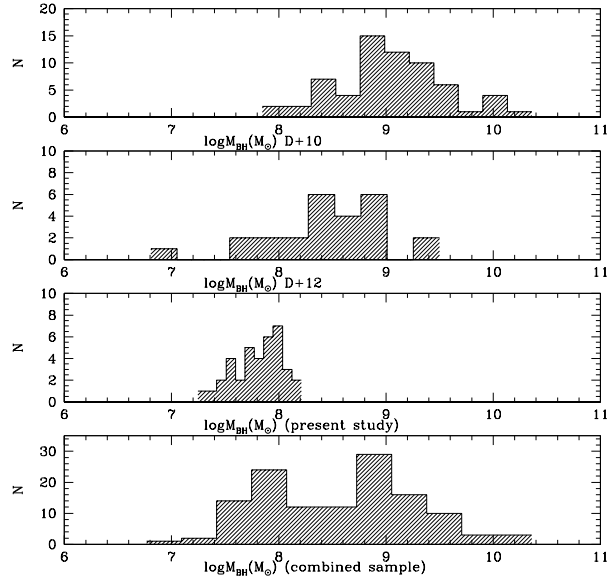}
\caption{Comparison of the M$_{BH}$ distribution of the quasars in \citet{dec10a} (top panel), \citet{decarli2012low} (second panel) and this study (third panel). The bottom panel shows the M$_{BH}$ distribution of the combined sample used in this study, emphasizing the enlarged, homogeneously covered parameter space.}
\label{a1}
\end{figure}

The quasars were selected using the following criteria: (1) classification as quasars in the SDSS database, based on optical spectra with S/N (Signal-to-noise ratio) $>$10; (2) virial black hole masses of $M_{BH}<10^{8.3}M_\odot$, calculated using the MgII emission line width and continuum luminosity from the SDSS spectra \citep{shen11}, following the method described in detail in  \citet{dec10a}; (3) the presence of at least two luminous stars within 1$^{\prime}$ projected distance from the quasar, so that the characterisation of the Point Spread Function (PSF) could be achieved with high enough precision and (4) the presence of at least two standard stars in the quasar field available in the 2MASS database for photometric calibration. About 200 quasars fulfilled these criteria, from which the final sample of 37 quasars were selected based on observability constraints and to cover as homogeneously  as possible the distribution of redshift, black hole mass and quasar luminosity. Figure~\ref{a1} compares the $M_{BH}$ distribution from \citet{dec10a}, \citet{decarli2012low} and this work, clearly showing the extension to lower BH masses afforded by the present study.\\

The black hole masses of the quasars, selected from the SDSS database, were estimated using the virial theorem. To estimate the size of the BLR, we have assumed the relations between the radius of the BLR and the AGN continuum luminosity, which are calibrated using reverberation mapping results for nearby AGN \citep[e.g.][]{kaspi05,grier12}. The continuum luminosity, together with the MgII emission line width contribute to provide the black hole mass estimate. The spectral analysis follows as performed in \citet{dec10a} and \citet{derosa11}. We model the AGN continuum with a power-law, a Balmer pseudo-continuum, and a template of the FeII emission \citep{vw01}. Since individual features of the FeII multiplets are typically heavily blended, we fix their width to the fitted MgII line width. The continuum is fitted in a wide spectral region around MgII (2500-2750 \AA{} and 2850-3000 \AA{}) and subtracted from the observed spectra. The MgII line is then fitted with two gaussian curves. The line width and the luminosities are measured from the best fit model. The continuum luminosity is interpolated at 3000 \AA{} over the best fit power-law component. The corresponding scale radius of the BLR is computed as $R_{\rm BLR}$/(10 light days) = $(2.52\pm0.30) [\lambda L_{\lambda}(3000\,\AA{}) / (10^{44} {\rm erg s^{-1}})]^{(0.47\pm0.05)}$.
Black hole masses are thus derived as:
\begin{equation}
M_{\rm BH} = \frac{R_{\rm BLR} (f \cdot {\rm FWHM})^2}{G}
\end{equation}
where, we assume $f$=1.6 as the geometrical factor accounting for our ignorance of the BLR dynamics and orientation. The value of $f$=1.6 is adopted in this work from \citet{dec08} derived for $\langle fH(\beta) \rangle$, assuming that MgII and H($\beta$) emitting regions share a similar geometry \citep{derosa11}.

\section{Imaging observations, data reduction and analysis}

The \textit{H}-band imaging observations of the quasars were obtained using the 2.5m Nordic Optical Telescope (NOT) on the Roque de los Muchachos, Spain, and the NOTCam instrument. A total of 37 quasars were observed during two observing runs in July 2012 and January 2013. With pixel scale of 0.234$^{\prime\prime}$/px, NOTCam provides a total field of view (FOV) of $\sim$4$^{\prime}\times4^{\prime}$. The average seeing during the observations was relatively good at $\sim$0.9$^{\prime\prime}$. Typical integration time was 40 min. per target in July 2012, and 63 min. per target in January 2013, as the quasars in that run were relatively fainter. A detailed summary of the observations is provided in Table 1.

\begin{table*}
\centerline{%
\begin{tabular}{lcccccccc}
\toprule 
Name & \textit{z} & m$_{\text{v}}$ & M$_{\text{v}}$ & $\log M_{BH}$  & Date & $\tau$ & FWHM & ZP\tabularnewline
    &     &     &     & ($M_{\odot}$) &   & (min.) & (arcsec) & \tabularnewline
(1) & (2) & (3) & (4) & (5) & (6) & (7) & (8) & (9)\tabularnewline
\midrule 
SDSS J013842.05+004020 & 0.520 & 19.0 & -23.4 & 7.82 & 23/01/2013 & 42 & 0.91 & 23.13\tabularnewline
SDSS J013912.8+152005  & 0.582 & 18.9 & -23.8 & 7.86 & 24/01/2013 & 63 & 0.82 & 23.02\tabularnewline
SDSS J023817.0-071810  & 0.605 & 19.3 & -23.5 & 8.21 & 24/01/2013 & 61 & 0.98 & 22.84\tabularnewline
SDSS J024141.52+000416 & 0.648 & 18.7 & -24.3 & 8.15 & 23/01/2013 & 60 & 0.85 & 23.10\tabularnewline
SDSS J074636.5+430206  & 0.513 & 18.9 & -23.5 & 7.98 & 23/01/2013 & 63 & 1.45 & 22.91\tabularnewline
SDSS J075517.5+152231  & 0.800 & 18.9 & -24.7 & 8.11 & 23/01/2013 & 60 & 1.04 & 22.93\tabularnewline
SDSS J080840.6+104738  & 0.714 & 19.0 & -24.3 & 8.04 & 24/01/2013 & 60 & 0.75 & 22.98\tabularnewline
SDSS J083437.0+532818  & 0.586 & 18.8 & -23.9 & 8.02 & 25/01/2013 & 63 & 0.80 & 22.71\tabularnewline
SDSS J091706.9+041723  & 0.745 & 19.0 & -24.4 & 7.93 & 23/01/2013 & 63 & 1.10 & 23.10\tabularnewline
SDSS J094838.3+184516  & 0.620 & 18.9 & -24.0 & 7.92 & 24/01/2013 & 61 & 0.91 & 22.91\tabularnewline
SDSS J102415.9+530019  & 0.892 & 19.3 & -24.6 & 8.07 & 23/01/2013 & 63 & 0.93 & 22.90\tabularnewline
SDSS J121020.7+521244  & 0.805 & 18.8 & -24.8 & 8.02 & 24/01/2013 & 63 & 0.82 & 23.00\tabularnewline
SDSS J124912.7+053014  & 0.679 & 18.8 & -24.3 & 7.92 & 24/01/2013 & 63 & 0.84 & 23.00\tabularnewline
SDSS J133401.2-024200  & 0.903 & 20.6 & -23.3 & 7.71 & 23/01/2013 & 61 & 1.16 & 23.01\tabularnewline
SDSS J134238.2+631347  & 0.539 & 20.5 & -22.0 & 7.57 & 25/01/2013 & 63 & 1.17 & 23.31\tabularnewline
SDSS J135128.15-001017 & 0.524 & 22.0 & -20.5 & 7.54 & 23/01/2013 & 34 & 0.68 & 23.03\tabularnewline
SDSS J135229.7+555034  & 0.638 & 20.5 & -22.5 & 7.69 & 25/01/2013 & 63 & 0.70 & 23.00\tabularnewline
SDSS J135501.3+015047 & 0.955 & 21.1 & -23.0 & 7.99 & 24/01/2013 & 61 & 0.71 & 23.00\tabularnewline
SDSS J132240.1+503011 & 0.780 & 19.9 & -23.6 & 7.83 & 01/07/2012 & 45 & 0.77 & 22.99\tabularnewline
SDSS J140917.5+220555 & 0.547 & 19.1 & -23.5 & 7.96 & 02/07/2012 & 41 & 0.66 & 22.29\tabularnewline
SDSS J145408.3+605517 & 0.940 & 20.4 & -23.6 & 7.71 & 03/07/2012 & 45 & 0.68 & 23.06\tabularnewline
SDSS J150518.4+022653 & 0.534 & 19.9 & -22.6 & 7.94 & 01/07/2012 & 44 & 0.95 & 22.94\tabularnewline
SDSS J155243.1+430605 & 0.860 & 20.4 & -23.4 & 8.00 & 02/07/2012 & 45 & 1.15 & 22.91\tabularnewline
SDSS J155858.2+232219 & 0.989 & 20.4 & -23.7 & 7.88 & 03/07/2012 & 45 & 0.63 & 22.71\tabularnewline
SDSS J160641.5+272557 & 0.543 & 20.9 & -21.7 & 7.29 & 02/07/2012 & 41 & 1.40 & 22.86\tabularnewline
SDSS J170134.2+254838 & 0.605 & 20.7 & -22.1 & 7.50 & 03/07/2012 & 37 & 0.72 & 22.83\tabularnewline
SDSS J170325.0+230001 & 0.566 & 20.4 & -22.3 & 7.75 & 02/07/2012 & 41 & 0.97 & 22.66\tabularnewline
SDSS J172357.0+541307 & 0.601 & 20.2 & -22.6 & 7.68 & 01/07/2012 & 45 & 0.72 & 22.93\tabularnewline
SDSS J173326.02+320813 & 0.713 & 20.2 & -23.1 & 7.66 & 01/07/2012 & 54 & 0.70 & 23.10\tabularnewline
SDSS J204443.3+010515  & 0.607 & 20.8 & -22.0 & 7.69 & 01/07/2012 & 48 & 0.92 & 22.75\tabularnewline
SDSS J204902.68+001803 & 0.510 & 20.7 & -21.7 & 7.84 & 03/07/2012 & 41 & 0.72 & 23.00\tabularnewline
SDSS J210114.99-003150 & 0.520 & 20.3 & -22.1 & 7.46 & 02/07/2012 & 41 & 0.90 & 22.90\tabularnewline
SDSS J210422.9-053650  & 0.646 & 19.2 & -23.8 & 7.98 & 01/07/2012 & 54 & 0.84 & 22.90\tabularnewline
SDSS J214138.53+000319 & 0.661 & 20.8 & -22.3 & 7.38 & 02/07/2012 & 41 & 0.97 & 22.90\tabularnewline
SDSS J220829.61-005024 & 0.750 & 20.7 & -22.7 & 7.57 & 03/07/2012 & 41 & 0.74 & 22.81\tabularnewline
SDSS J223925.9+000341  & 0.586 & 19.7 & -23.1 & 7.57 & 02/07/2012 & 41 & 1.65 & 22.90\tabularnewline
SDSS J234227.39-000125 & 0.863 & 19.4 & -24.4 & 7.83 & 03/07/2012 & 36 & 0.94 & 22.50\tabularnewline
\bottomrule
\end{tabular}}
\caption{{\small The properties of the sample and journal of observations. 
(1) Target name. (2) Redshift. (3) The apparent
V-band magnitude. (4) The absolute V-band magnitude. 
(5) Estimated  virial black hole mass. (6) The date of imaging observation. 
(7) Total integration time. (8) Seeing FWHM. 
(9) Photometric zero-point from standard stars. }}
\end{table*}

\subsection{Data Reduction}

Data reduction was performed using the NOTCam quicklook package based on IRAF scripts \footnote{IRAF is distributed by the National Optical Astronomy Observatories, which are operated by the Association of Universities for Research in Astronomy, Inc., under cooperative agreement with the National Science Foundation}. A mask file obtained from the NOTCam bad pixel mask archive was employed to mask the bad pixels in the image frames.  Flat fielding was performed using a normalised median combined masterflat obtained from a pair of sky flats. For sky subtraction, a scaled sky template was produced from a list of dithered flat-fielded image frames. Finally, the flat-fielded, sky-subtracted images were aligned to within sub-pixel accuracy and combined to obtain the final reduced co-added image. Zero point calibration was performed by cross-matching the photometry of field stars with the 2MASS database in the \textit{H}-band.

\subsection{2D image analysis}

To derive the properties of the quasar host galaxies, we have used 2D model fitting of the surface brightness distribution, assuming that the image of the quasar is a superposition of the nucleus and the surrounding nebulosity. To perform this 2D data analysis, an IDL 6.0 based software package called AIDA (Astronomical Image Decomposition and Analysis)\citep{uf11} was employed. The nucleus is described by the local PSF of the image and the host galaxy is modelled by a Sersic law convolved with the PSF. Our analysis carefully follows similar modelling strategy as described in detail in \citet{decarli2012low}.

\subsubsection{PSF modelling}

The modelling of the PSF is crucial to estimate the emission from the nuclear source against which the extended light from the host galaxy will be observed. In order to model the PSF, we employed multiple field stars within the frame of each quasar, that were bright enough to model the faint PSF wing. For the selection of the field stars, we preferentially chose stars with the smallest FWHM and ellipticity and having brightest magnitude in the field. Contamination near the regions close to the stars, including companions, saturated pixels or other defects were carefully masked out. The selected stars were each modelled with four 2D gaussians and an exponential feature. The gaussians and the exponential feature represented the core of the PSF and the extended PSF wings, respectively. A circular annulus centered around the stars was used to calculate the local background. The final PSF fit region was selected after defining the internal and external radius of the circular area. This allowed us to remove the cores of bright, saturated stars. Typical uncertainty in the PSF model is $\leq\sim$0.3 mag. For a detailed discussion on the PSF characterisation, see \citet{jari07,jari09}.   

\subsubsection{Host characterisation}

\begin{figure}
\centering
\begingroup
\sbox0{\includegraphics[scale=0.54]{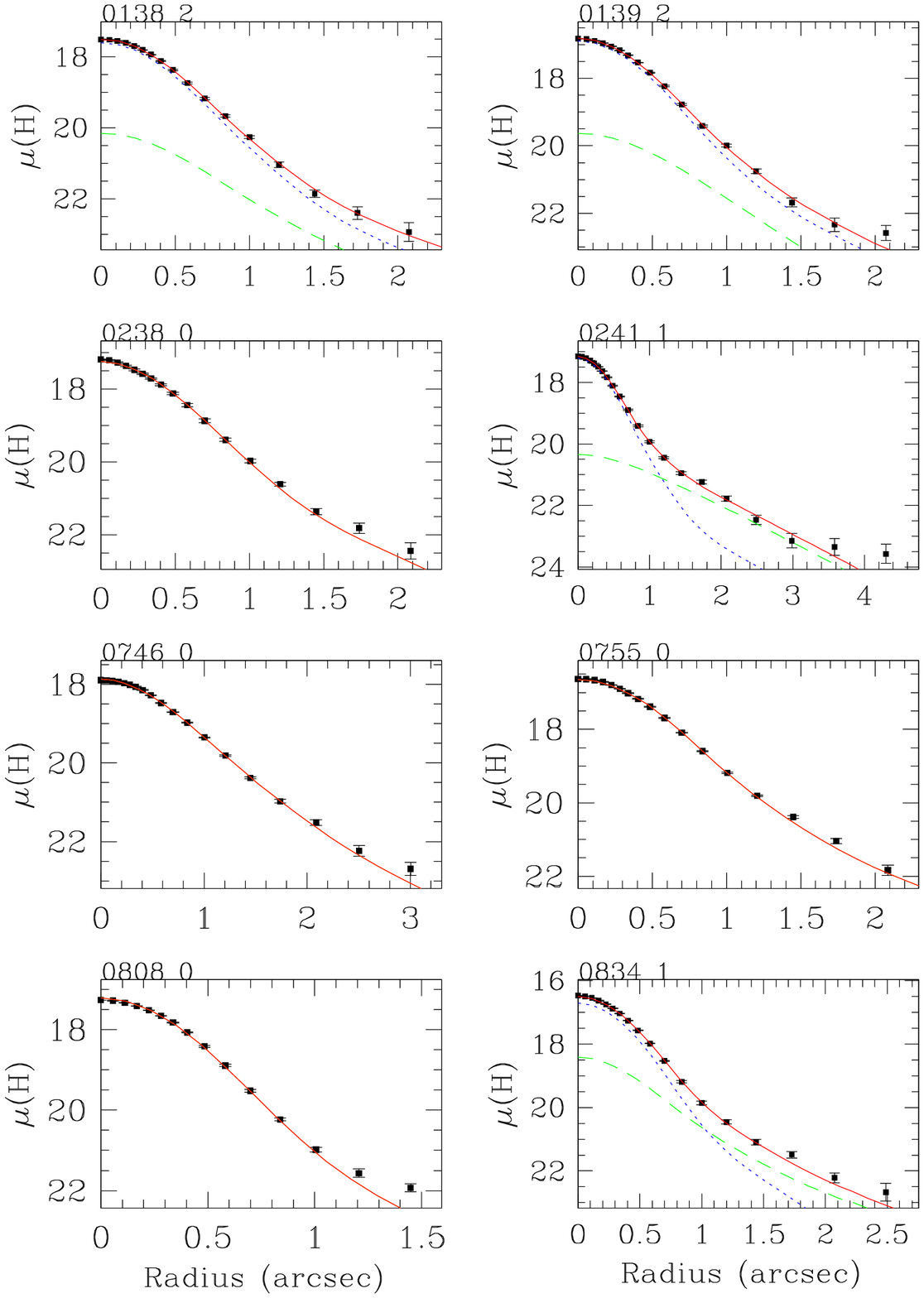}}%
\includegraphics[clip,trim={0.79\wd0} {0.81\wd0} 0 0]{profile2.eps}
\endgroup
\caption{Profiles of H-band radial surface brightness versus radius in arcsecs, are shown for 3 QSOs (from top to bottom - SDSS J013912.8+152005, SDSS J024141.52+000416, SDSS J075517.5+152231), overlaid with scaled PSF model (blue, dotted line), the de Vaucouleurs r$^{1/4}$ model convolved with the PSF (green, long dashed line) and the fitted PSF+host galaxy model profile (red, solid line). From top to bottom, they are classified as Marginally resolved, Resolved and Unresolved case. For a complete list of such profiles of all the quasars in this study, please refer the online electronic version of this paper.}
\label{a2}
\end{figure}

After modelling the PSF, in order to distinguish between the resolved and unresolved quasars, the images of the quasars were first fitted with a pure PSF model. If there was extended emission observed in the residuals or if the residuals showed a significant excess over the PSF shape, in a region relatively far away from the nucleus (typically between 1$^{\prime\prime}$-2$^{\prime\prime}$), then we considered the quasar to be resolved. The resolved quasars were further fitted with a two component (PSF + galaxy) model. The fitting was performed using the Sersic 
model convolved with the PSF to describe the host galaxy and a scaled PSF to describe the nucleus. To evaluate whether the quasar is resolved, marginally resolved or unresolved, we investigated the $\chi^2$ ratio of the fits (see Table 2), statistically providing the amount of deviation of the PSF + galaxy model from the pure PSF fit. For a resolved quasar, the PSF + galaxy model is expected to be better than the pure PSF fit. In most cases, the ratio is above 1.1 for resolved quasars, around 1 for unresolved quasars and between these values for the marginally resolved quasars. In some exceptional cases, the values represented by the $\chi^2$ ratio are not sufficient to characterise whether the quasar is resolved or not. Therefore, we conducted additional visual inspection of the profiles, examples of which are shown in Figure~\ref{a2}, to robustly ascertain whether the PSF + galaxy fit was better than the pure PSF fit. For the unresolved quasars, we visually determined the upper limit to their host galaxy luminosity by adding increasingly bright galaxy models to the PSF model, upto the point where the model profile became inconsistent with the observed profile within the observational error bars. There are four quasars which fit neither the resolved, nor the unresolved class, and we classify them as marginally resolved cases. An example of the output of this analysis is shown in Figure~\ref{a2}. For a detailed discussion on the host galaxy characterisation, see \citet{jari07,jari09}.

In Table 2, it can be seen that the Sersic index $n_s$ for a large fraction of the quasars tends to be either 0.90 or 5.00, the extreme values considered by the AIDA pipeline. However, Table 2 shows also that 50\% of the marginally resolved quasars have well-defined (i.e. not extreme) Sersic index values, indicating that there are no systematic errors. We also note that $n_s$ does not show inverse correlation with the effective radius, as would be expected if the pipeline were erroneously identifying a PSF with broad wings as a small bulge. 

The 2D model fitting with AIDA provides us with the apparent host galaxy magnitudes in the \textit{H}-band. These apparent magnitudes were converted to the rest frame \textit{R}-band absolute magnitudes, using the elliptical galaxy template from \citep{man01} to estimate the required \textit{K}-correction. Then the \textit{R}-band absolute magnitudes of the host galaxies were converted into stellar mass, assuming the mass-to-light(M/L) ratio of a single stellar population originating at \textit{z}$_{burst}$=5 and passively evolving down to \textit{z}=0 \citep{dec10b}. 

\subsubsection{Error Estimation}

The errors associated to the parameters of the best fit depend on a number of assumptions based on the effective observing conditions (e.g. for the residual uncertainty of the background subtraction which particularly affects near-IR images) that are not always well understood. Moreover, a number of systematic effects can further introduce uncertainties. The best way to estimate the uncertainties of the fitting parameters, is to produce simulated images of the objects to try to achieve similar parameters and conditions, as those from real observations and thereby to evaluate the uncertainty of a given parameter by analysing multiple mock observations. In order to perform such a test for each object, we produced simulated images of the targets assuming the best fit parameters and the noise level derived from the real observations. The instrumental setup was fixed by the values of pixel scale, read-out noise, gain, exposure time, number of combined exposures and the zero point of the calibration. The simulated images were produced using the AETC tool\footnote{AETC:Advanced Exposure Time Calculator: http://aetc.oapd.inaf.it/} and then analysed with AIDA using the same conditions adopted for the analysis of the real images but not using information about the true parameters of the target. For each target, we performed 25 simulations and then compared the distribution of the parameters with those derived from the best fit of the real data. It turned out that the total magnitude of the objects always lies within a few millimagnitudes, whereas for the host galaxy magnitude the typical uncertainty ranges from 0.2 magnitude for well resolved sources to 0.6-0.8 magnitude for the marginally resolved sources. We also report that the uncertainties in the effective radius for well resolved objects are within 20-30\% in most cases, while for marginally resolved objects the uncertainty can be as high as 80\%. The most critical parameter is the index $n_s$ of the Sersic model.  

\section{Results and Discussions}

\begin{figure}
\includegraphics[height=9cm]{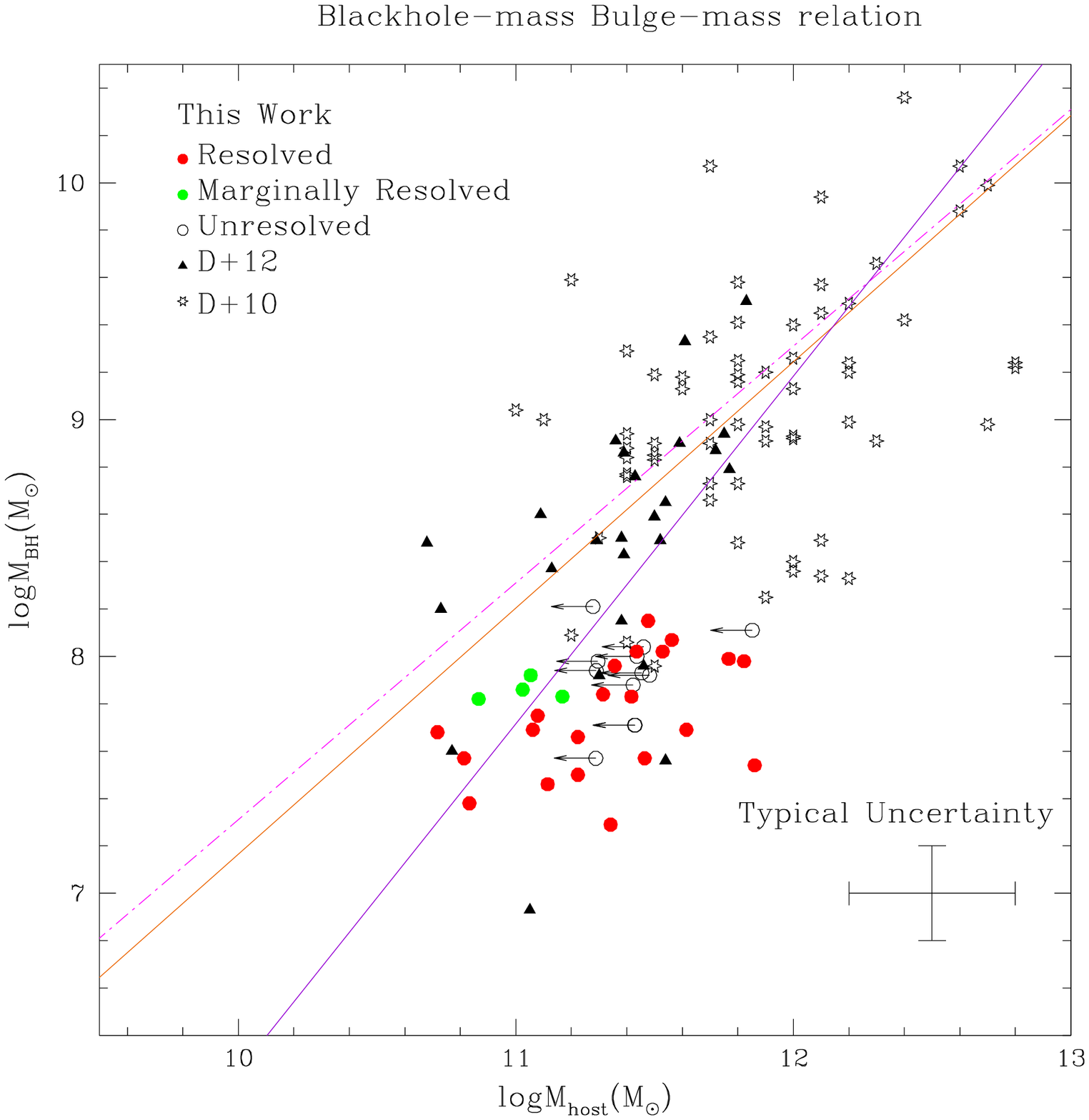}
\caption{The $M_{BH}$-$M_{host}$ relation for quasars in this work (red circles: resolved, green circles: marginally resolved, and open circles: unresolved), \citet{dec10a} and \citet{dec10b} (open stars) and \citet{decarli2012low} (black triangles). The best bilinear regression fit is shown for the entire resolved sample (violet solid line) and for the quasars with log$M_{BH}>8.2(M_\odot)$(dark orange solid line). The regression fit for the local relation for inactive galaxies by \citet{marconi_hunt} is shown as magenta dashed line, with $\langle\Gamma\rangle\sim0.002$. A typical error bar is shown in the lower right corner.}
\label{a4}
\end{figure}

Out of our sample of 37 quasars with low mass black holes, 21 (57\%) are resolved, 4 (11\%) are marginally-resolved and 12 (32\%) remain unresolved. We have produced the full quasar sample with BH masses up to $M_{BH}\leq10^{10}M_\odot$ by including all the quasars at z $<$ 1 from Decarli et al.(2010a,b; 2012). The $M_{BH}$-$M_{host}$ relation up to \textit{z}$<$1.0 is shown in Figure~\ref{a4}. As can be seen, the full quasar sample defines a linear relation. However, the quasars with $M_{BH}<10^8M_\odot$ deviate from the relation defined by the quasars at $M_{BH}>10^{8.2}M_\odot$ and the local relation \citep{marconi_hunt}. The $M_{BH}$-$M_{host}$ relation becomes indeed steeper and slightly tighter at the lowest BH masses, causing the linear relation to apparently break down at $M_{BH}\sim10^{8.2}M_\odot$.
The best bilinear regression fits of the entire sample of resolved quasars (eq. no. 2) and for resolved quasars with $M_{BH}>10^{8.2}M_\odot$ (eq. no. 3) are given as:
\begin{equation}
log\frac{M_{\rm BH}}{10^{8.5}M_{\odot}} = (1.46\pm0.25) \times log\frac{M_{\rm host}}{10^{11.7}M_{\odot}} + (0.24\pm0.11)
\end{equation}
\begin{equation}
log\frac{M_{\rm BH}}{10^{8.5}M_{\odot}} = (1.03\pm0.24) \times log\frac{M_{\rm host}}{10^{11.7}M_{\odot}} + (0.43\pm0.10)
\end{equation}

Such a break in the relation could in principle be caused by a systematic a) underestimation of $M_{BH}$ and/or b) overestimation of $M_{host}$, at low BH masses. We believe that neither explanation is likely, as a) BH masses based on SDSS spectra have been shown to be robust in numerous previous studies \citep[e.g.][]{rh11,labita09}, and the amount of uncertainty required to explain the break is much higher than the expected uncertainty in deriving $M_{BH}$. As for case b), the majority of our sample quasars are well resolved, leading to a robust determination of the host galaxy luminosity/mass \citep[e.g.][]{falomo2014}. In addition, the amount of systematic error would require an order of magnitude underestimation of the PSF model. Hence, the most likely reason for the apparent break in the $M_{BH}$-$M_{host}$ relation below $M_{BH}\sim10^{8.2}M_\odot$ is due to significant disc contamination. 
\\

Disc correction was therefore performed for quasars having disc components because bulge mass is better correlated with $M_{BH}$ than the disc mass or the total host galaxy mass. Due to the seeing-limited spatial resolution of ground-based observations, accurate decomposition of quasar host galaxies at high-\textit{z} into bulge and disc components is practically impossible. However, the Sersic index 
$n_s$ can be used to estimate the galaxy morphology such that, $n_s\sim1$ indicates disc-dominated galaxies whereas, $n_s\sim4$ indicates elliptical galaxies. \citet{sim11} performed bulge-disc decomposition of $\sim$10$^6$ galaxies from the SDSS by analytical estimation. \citet{decarli2012low} followed similar analytical approach which is also adopted in this work, for correcting the disc component.
\begin{equation}
\frac{\rm B}{\rm T}=\frac{n_s-0.5}{3.5} 
\end{equation}
for $n_s<4$ 

\begin{figure}
\includegraphics[height=9cm,angle=0]{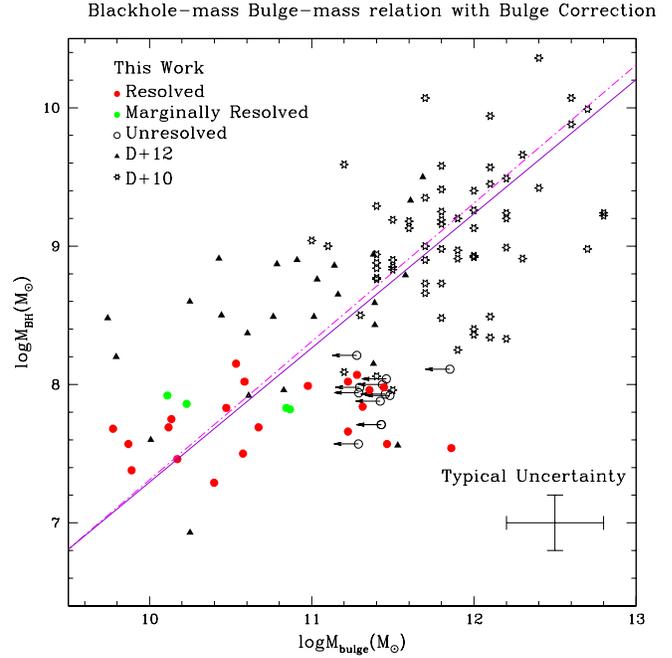}
\caption{The $M_{BH}$-$M_{bulge}$ relation after the disc correction.The best bilinear regression fit is shown for the entire resolved sample as violet solid line while the magenta dashed line shows the regression fit for the local relation for inactive galaxies by \citet{marconi_hunt} with $\langle\Gamma\rangle\sim0.002$. For meaning of symbols, see Figure~\ref{a4} caption.}
\label{a5}
\end{figure}

The above analytical equation for disc correction can be used to estimate a range of $\frac{\rm B}{\rm T}$ values for $n_s<4$. No disc correction is required for $n_s\geq4$, i.e. pure spheroidal galaxies. Figure~\ref{a5} shows the $M_{BH}$ - $M_{bulge}$ relation after performing the disc correction. A detailed summary of our results is shown in Table-2. The best bilinear regression fit of the $M_{BH}$-$M_{bulge}$ relation (after disc correction), considering all the resolved quasar samples is given as:
\begin{equation}
log\frac{M_{\rm BH}}{10^{8.5}M_{\odot}} = (0.97\pm0.15) \times log\frac{M_{\rm host}}{10^{11.7}M_{\odot}} + (0.44\pm0.11)
\end{equation}

The slope of the relation becomes indeed flatter than the slope of the relation without disc correction. After correcting for the disc contamination, the best fit of the relation is in very good agreement with the local relation, indicating that the apparent break in the regression fit is caused by disc dominated galaxies. 

It is evident from Fig. 4 that the quasars from \citet{decarli2012low} corrected for the disc contamination are in similar mass scales in $M_{bulge}$ as our disc corrected quasars. The overall scatter of the relation does not change significantly after the disc correction. Consecutively, $\langle log\Gamma \rangle$($\equiv\frac{M_{BH}}{M_{host}}$) increases from -2.98$\pm$0.0511 to -2.74$\pm$0.0536. It is worth to note that this study extends the parameter space of the disc-corrected relation up to $\sim$3.5 dex in both $M_{BH}$ and $M_{bulge}$ towards the low-mass end.

Galaxies can be classified based on their surface brightness profiles as core-Sersic galaxies (typically massive and luminous having undergone dry mergers) and Sersic galaxies (typically less massive; $M_{host} \leq 3\times 10^{10} M_\odot$). \citet{scott13} found the best bilinear regression fits for nearby core-Sersic and Sersic galaxies as $M_{BH}$ $\propto$ $M_{host}^{0.80}$ and M$_{BH}$ $\propto$ $M_{host}^{1.80}$ respectively. This study includes eight resolved quasars that can be classified as Sersic galaxies. Furthermore, 62 quasars from \citet{dec10a} and \citet{dec10b} and 18 quasars from \citet{decarli2012low} can be classified as purely core-Sersic galaxies and three quasars as purely Sersic galaxies. It is worth to mention that for our quasars with disc dominated host galaxies, their bulge masses are considered. The bilinear regression for core-Sersic and Sersic galaxies in this study up to z$<$1 extending into the low mass region are given as M$_{BH}$ $\propto$ $M_{host}^{0.80}$ and M$_{BH}$ $\propto$ $M_{host}^{1.22}$ respectively.
\\
\\
\pagebreak
\begin{sidewaystable}
\noindent\resizebox{\textwidth}{!}{
\begin{tabular}{lcccccrcrcrcc}
\toprule 
Name & \textit{z} & $\chi^2_{psf}$/$\chi^2_{gq}$ & Note & n$_{sersic}$ & m$_{nuc}$ & m$_{host}$ & $k-corr$ & M$_{R}$ & log(M/L) & LogM$_{host}$ & B/T & LogM$_{bulge}$\tabularnewline
 & & & & & & & & & & (M$_{\odot}$) & & (M$_{\odot}$)\tabularnewline
(1) & (2) & (3) & (4) & (5) & (6) & (7) & (8) & (9) & (10) & (11) & (12) & (13)\tabularnewline
\midrule
SDSS J013842.05+004020 & 0.520 & 1.02 & M & 5.00 & 17.44 & 18.94 & 2.56 & -20.85 & 0.62 & 10.86 & -- & --\tabularnewline
SDSS J013912.8+152005 & 0.582 & 1.09 & M & 1.06 & 16.89 & 18.81 & 2.54 & -21.30 & 0.60 & 11.02 & 0.160 & 10.22\tabularnewline
SDSS J023817.0-071810 & 0.605 & 1.00 & U & -- & 16.94 & $>$18.26 & 2.54 & $>$-21.96 & 0.59 & $>$11.27 & -- & --\tabularnewline
SDSS J024141.52+000416 & 0.648 & 1.66 & R & 0.90 & 17.20 & 17.93 & 2.53 & -22.48 & 0.58 & 11.47 & 0.114 & 10.53\tabularnewline
SDSS J074636.5+430206 & 0.513 & 1.01 & U & -- & 16.70 & $>$17.82 & 2.57 & $>$-21.93 & 0.62 & $>$11.29 & -- & --\tabularnewline
SDSS J075517.5+152231 & 0.800 & 1.01 & U & -- & 16.19 & $>$17.50 & 2.48 & $>$-23.52 & 0.54 & $>$11.85 & -- & --\tabularnewline
SDSS J080840.6+104738 & 0.714 & 1.01 & U & -- & 17.54 & $>$18.19 & 2.51 & $>$-22.49 & 0.56 & $>$11.46 & -- & --\tabularnewline
SDSS J083437.0+532818 & 0.586 & 1.52 & R & 2.65 & 16.93 & 17.80 & 2.54 & -22.32 & 0.60 & 11.43 & 0.614 & 11.22\tabularnewline
SDSS J091706.9+041723 & 0.745 & 1.00 & U & -- & 18.51 & $>$18.31 & 2.50 & $>$-22.50 & 0.55 & $>$11.45 & -- & --\tabularnewline
SDSS J094838.3+184516 & 0.620 & 1.00 & U & -- & 16.54 & $>$17.82 & 2.53 & $>$-22.47 & 0.59 & $>$11.48 & -- & --\tabularnewline
SDSS J102415.9+530019 & 0.892 & 1.10 & R & 2.33 & 17.35 & 18.46 & 2.46 & -22.86 & 0.51 & 11.56 & 0.522 & 11.27\tabularnewline
SDSS J121020.7+521244 & 0.805 & 1.40 & R & 0.90 & 17.04 & 18.30 & 2.48 & -22.73 & 0.53 & 11.52 & 0.114 & 10.58\tabularnewline
SDSS J124912.7+053014 & 0.679 & 1.05 & M & 0.90 & 16.78 & 19.09 & 2.52 & -21.44 & 0.57 & 11.05 & 0.114 & 10.11\tabularnewline
SDSS J132240.1+503011 & 0.780 & 1.02 & M & 2.16 & 18.42 & 19.13 & 2.49 & -21.80 & 0.54 & 11.16 & 0.474 & 10.84\tabularnewline
SDSS J133401.2-024200 & 0.903 & 0.89 & U & -- & 18.63 & $>$18.82 & 2.46 & $>$-22.54 & 0.51 & $>$11.43 & -- & --\tabularnewline
SDSS J134238.2+631347 & 0.539 & 1.16 & R & 0.90 & 18.75 & 19.15 & 2.56 & -20.74 & 0.61 & 10.81 & 0.114 & 9.86\tabularnewline
SDSS J135128.15-001017 & 0.524 & 4.91 & R & 4.16 & 19.79 & 16.48 & 2.56 & -23.34 & 0.62 & 11.86 & -- & --\tabularnewline
SDSS J135229.7+555034 & 0.638 & 1.13 & R & 0.90 & 18.34 & 18.95 & 2.53 & -21.41 & 0.59 & 11.06 & 0.114 & 10.11\tabularnewline
SDSS J135501.3+015047 & 0.955 & 1.67 & R & 1.07 & 19.42 & 18.12 & 2.44 & -23.40 & 0.50 & 11.76 & 0.162 & 10.97\tabularnewline
SDSS J140917.5+220555 & 0.547 & 1.15 & R & 5.00 & 16.42 & 17.83 & 2.56 & -22.10 & 0.61 & 11.35 & -- & --\tabularnewline
SDSS J145408.3+605517 & 0.940 & 1.02 & U & -- & 18.06 & $>$18.93 & 2.44 & $>$-22.56 & 0.50 & $>$11.42 & -- & --\tabularnewline
SDSS J150518.4+022653 & 0.534 & 1.01 & U & -- & 17.29 & $>$17.93 & 2.56 & $>$-21.94 & 0.61 & $>$11.29 & -- & --\tabularnewline
SDSS J155243.1+430605 & 0.860 & 1.00 & U & -- & 17.92 & $>$18.69 & 2.47 & $>$-22.53 & 0.52 & $>$11.43 & -- & --\tabularnewline
SDSS J155858.2+232219 & 0.989 & 1.01 & U & -- & 18.07 & $>$19.07 & 2.43 & $>$-22.57 & 0.49 & $>$11.42 & -- & --\tabularnewline
SDSS J160641.5+272557 & 0.543 & 1.15 & R & 0.90 & 17.70 & 17.84 & 2.56 & -22.06 & 0.61 & 11.34 & 0.114 & 10.39\tabularnewline
SDSS J170134.2+254838 & 0.605 & 1.13 & R & 1.29 & 18.77 & 18.39 & 2.54 & -21.82 & 0.59 & 11.22 & 0.225 & 10.57\tabularnewline
SDSS J170325.0+230001 & 0.566 & 1.04 & R & 0.90 & 18.08 & 18.60 & 2.55 & -21.43 & 0.60 & 11.07 & 0.114 & 10.13\tabularnewline
SDSS J172357.0+541307 & 0.601 & 1.03 & R & 0.90 & 18.15 & 19.66 & 2.54 & -20.53 & 0.60 & 10.71 & 0.114 & 9.77\tabularnewline
SDSS J173326.02+320813 & 0.713 & 1.06 & R & 5.00 & 18.52 & 18.78 & 2.51 & -21.89 & 0.56 & 11.22 & -- & --\tabularnewline
SDSS J204443.3+010515 & 0.607 & 1.64 & R & 0.90 & 19.10 & 17.43 & 2.53 & -22.80 & 0.59 & 11.61 & 0.114 & 10.67\tabularnewline
SDSS J204902.68+001803 & 0.510 & 1.42 & R & 5.00 & 19.10 & 17.76 & 2.57 & -21.97 & 0.62 & 11.31 & -- & --\tabularnewline
SDSS J210114.99-003150 & 0.520 & 1.24 & R & 0.90 & 18.64 & 18.32 & 2.56 & -21.47 & 0.62 & 11.11 & 0.114 & 10.17\tabularnewline
SDSS J210422.9-053650 & 0.646 & 2.86 & R & 1.97 & 18.37 & 17.05 & 2.53 & -23.34 & 0.58 & 11.82 & 0.420 & 11.44\tabularnewline
SDSS J214138.53+000319 & 0.661 & 1.02 & R & 0.90 & 19.71 & 19.60 & 2.52 & -20.87 & 0.58 & 10.83 & 0.114 & 9.88\tabularnewline
SDSS J220829.61-005024 & 0.750 & 1.13 & R & 5.00 & 18.30 & 18.30 & 2.50 & -22.52 & 0.55 & 11.46 & -- & --\tabularnewline
SDSS J223925.9+000341 & 0.586 & 1.03 & U & -- & 17.23 & $>$18.17 & 2.54 & $>$-21.96 & 0.60 & $>$11.28 & -- & --\tabularnewline
SDSS J234227.39-000125 & 0.863 & 1.04 & R & 0.90 & 17.78 & 18.75 & 2.47 & -22.48 & 0.52 & 11.41 & 0.114 & 10.47\tabularnewline
\bottomrule
\end{tabular}}
\caption{{Results from the image analysis. 
(1) Quasar name. (2) Redshift. (3) Ratio of $\chi^2$ values between the best fit using the pure PSF model and the best fit using a PSF + galaxy model. (4) Note specifying whether the target is resolved (R), marginally resolved (M) or unresolved (U). (5) Sersic index of the host galaxy model. (6) Apparent observed H-band magnitude of the nucleus. (7) Apparent observed H-band magnitude of the host galaxy. (8) K-correction between the observed H-band and the rest frame R-band. (9) Resulting absolute rest-frame R-band magnitude of the host galaxy. (10) Adopted mass-to-light ratio. (11) Stellar mass of the host galaxy. (12) Adopted bulge luminosity to total host luminosity ratio. (13) Stellar mass of the host galaxy bulge after the disc correction.}}
\end{sidewaystable}
\\
\pagebreak

In accordance with the morphological analysis and the bulge + disc decomposition performed in this work, we find that $\sim$3/4 of all the resolved and marginally resolved quasars possess significant disc components. Since most of these quasars lie below the local relation, we can infer that, for these bulges to increase and fit themselves with the local relation, a mechanism is required that would redistribute the stellar content of the galaxy. Such a scenario has already been encountered in the previous low-mass quasar studies, e.g. \citet{schramm_silvermann} and \citet{decarli2012low}. This makes it vital to correct the host galaxy masses for the disc component, to consider the bulge component only. Recently, \citet{kor11} and \citet{sani11} have suggested that galaxies that lie below the log-linear $M_{BH}$-$M_{host}$ mass relation of inactive galaxies, are evidently pseudobulges. On the contrary, \citet{g12} argues that galaxies hosting low-mass SMBHs are not pseudobulges, but rather an overlooked case of non-log-linear nature of $M_{BH}$-$M_{host}$ relation for classical spheroids. Although the relation for the Sersic galaxies in our study is steeper than for the core-Sersic galaxies, as seen in the previous discussion, we do not find evidence for a "break" in the slope, as advocated by \citet{g12} and \citet{scott13}, as we find that a single log-linear relation is recovered after the disc correction. The secular evolution of galaxy discs can allow the stars and gas within the galaxy to redistribute themselves in response to instabilities. In that case, mergers are not a likely scenario responsible for the stellar redistribution in the discs. Similarly, \citet{jiang} found for a sample of 147 nearby active galaxies with $M_{BH}$ ranging between 10$^5M_{\odot}$-10$^6M_{\odot}$ that a higher fraction of galaxies were disc-dominated likely containing pseudobulges, evolving secularly. Our results are consistent with those by \citet{kor11} and \citet{kor112}, who found that high-mass pseudobulges with $M_{BH} > 10^{7} M_\odot$ follow the $M_{BH}$- $M_{host}$ correlation. Only at lower masses, which our quasar samples do not probe, the pseudobulges start to deviate from the BH-host correlation.

\citet{dec10b} found evidence for strong evolution in $\Gamma$ in their large sample of high mass quasars, increasing by a factor $\sim$2 from \textit{z}=0 to \textit{z}=1 (log$\Gamma$ increases from -2.91 to -2.63). At lower masses, we find that the $\langle log\Gamma \rangle$, considering the bulge component (i.e. disc-corrected galaxy masses), evolves from log$\Gamma\sim-2.39\pm0.12$ at $\langle \textit{z} \rangle$=0.286 in \citet{decarli2012low} to log$\Gamma\sim-2.98\pm0.12$ at $\langle \textit{z} \rangle$=0.653 in this work, i.e. a decrease in $\Gamma$ by a factor $\sim$3.5. There is, however, a marked difference in the $M_{BH}$ distribution of the samples, so it is not possible to ascertain how much of the observed evolution, if any, is due to real redshift evolution.

\section{Conclusions}

The $M_{BH}$ and host galaxy properties of a sample of 37 low-mass quasars were combined with 89 quasars from the literature up to \textit{z} $<$ 1.0 and up to $M_{BH}<10^{11}M_\odot$, to investigate the relationship between BH-mass ($M_{BH}$) and host galaxy mass ($M_{host}$). From this study we conclude that:\\
\\
1. There is an apparent break at $M_{BH}\sim10^{8.2}M_\odot$ in the $M_{BH}$- $M_{host}$ relation.\\
2. The host galaxies of $\sim$75\% of the quasars in our sample of low-mass quasars, is dominated by a disc component.\\
\\
3. After correcting for the disc component, the linear $M_{BH}$- $M_{bulge}$ relation holds over the entire parameter space [(10$^{6.9}$ - 10$^{10.4}M_{BH}(M_\odot)$)$\sim$3.5 dex for M$_{BH}$ and (10$^{9.5}$ - 10$^{13}M_{bulge}(M_\odot)$)$\sim$3.5 dex for M$_{bulge}$].\\
\\
4. The $M_{BH}$- $M_{bulge}$ relation is in very good agreement with the local relation after disc correction.\\
\\
5. We discuss the scenario of secular evolution of discs of galaxies being responsible for the relatively strong disc domination.\\

|\section*{Acknowledgments}|
Based on observations made with the Nordic Optical Telescope, operated by the Nordic Optical Telescope Scientific Association at the Observatorio del Roque de los Muchachos, La Palma, Spain, of the Instituto de Astrofisica de Canarias.

\label{lastpage}

\end{document}